\documentclass[preprint]{aastex63}
\usepackage{graphicx}
\usepackage{verbatim}
\usepackage{tikz}

\usepackage{placeins}

\usepackage{amsmath}

\definecolor{forestgreen}{rgb}{0.10, 0.50, 0.10}

\received{\today}
\revised{ xxx }
\accepted{ xxx }
\submitjournal{xxx}

\shorttitle{Quiet Sun imaging of 20-80~MHz}
\shortauthors{Zhang et al.}
\begin{document}
%\begin{linenumbers}

%\title{Parametric simulation studies on the wave propagation of solar radio emission:\\ Angular distribution, brightness temperature, and shape (eccentricity)}

\title{Imaging of the Quiet Sun in the Frequency Range of 20-80~MHz}

\correspondingauthor{Pietro Zucca}
\email{zucca@astron.nl}

%\author[0000-0001-6855-5799]{PeiJin Zhang}

\author[0000-0001-6855-5799]{PeiJin Zhang}

\affiliation{Institute of Astronomy and National Astronomical Observatory,\\ Bulgarian Academy of Sciences, Sofia 1784, Bulgaria}
\affiliation{ASTRON, The Netherlands Institute for Radio Astronomy,\\
		Oude Hoogeveensedijk 4, 7991 PD Dwingeloo, The Netherlands}
\affiliation{Astronomy \& Astrophysics Section, Dublin Institute for Advanced Studies, Dublin 2, Ireland. }	
\affiliation{CAS Key Laboratory of Geospace Environment,
	School of Earth and Space Sciences, \\
	University of Science and Technology of China,
	Hefei 230026, China}

\author[0000-0002-6760-797X]{Pietro Zucca$^*$}
\affiliation{ASTRON, The Netherlands Institute for Radio Astronomy,\\
		Oude Hoogeveensedijk 4, 7991 PD Dwingeloo, The Netherlands}

\author[0000-0002-6591-4482]{Kamen Kozarev}
\affiliation{Institute of Astronomy and National Astronomical Observatory,\\ Bulgarian Academy of Sciences, Sofia 1784, Bulgaria}

\author[0000-0002-6106-5292]{Eoin Carley}
\affiliation{Astronomy \& Astrophysics Section, Dublin Institute for Advanced Studies, Dublin , Ireland. }	

\author[0000-0001-6252-5580]{ChuanBing Wang}
\affiliation{CAS Key Laboratory of Geospace Environment,
	 School of Earth and Space Sciences, \\
	 University of Science and Technology of China,
	  Hefei 230026, China}
\affiliation{CAS Center for the Excellence in Comparative Planetology, 
    \\University of Science and Technology of China,  Hefei 230026, China}
\affiliation{Mengcheng National Geophysical Observatory, School of Earth and Space Sciences, University of Science and Technology of China, Hefei 230026, China}

\author[0000-0001-6449-9611]{Thomas Franzen}
\affiliation{ASTRON, The Netherlands Institute for Radio Astronomy,\\
		Oude Hoogeveensedijk 4, 7991 PD Dwingeloo, The Netherlands}

\author[0000-0002-4705-7798]{Bartosz Dabrowski}
\affiliation{Space Radio-Diagnostics Research Centre, University of Warmia and\\
Mazury, R. Prawocheńskiego 9, 10-719 Olsztyn, Poland }	

\author[0000-0003-2812-6222]{Andrzej Krankowski}
\affiliation{Space Radio-Diagnostics Research Centre, University of Warmia and\\
Mazury, R. Prawocheńskiego 9, 10-719 Olsztyn, Poland }	

\author[0000-0003-1169-3722]{Jasmina Magdalenic}
\affiliation{
SIDC, Royal Observatory of Belgium, Brussels, Belgium and CmPA/Dept. of Mathematics, KU Leuven, Leuven, Belgium}

\author[0000-0001-8583-8619]{Christian Vocks}
\affiliation{Leibniz-Institut f{\"u}r Astrophysik Potsdam: Potsdam, Brandenburg, Germany}

%\author{the LOFAR Solar and Space Weather KSP Team}



\begin{abstract}
Radio emission of the quiet Sun is considered to be due to thermal bremsstrahlung emission of the hot solar atmosphere. The properties of the quiet Sun in the microwave band have been well studied, and they can be well described by the spectrum of bremsstrahlung emission. In the meter-wave and decameter-wave bands, properties of the quiet Sun have rarely been studied due to the instrumental limitations. In this work, we use the \textit{LOw Frequency ARray} (LOFAR) telescope to perform high quality interferometric imaging spectroscopy observations of quiet Sun coronal emission at frequencies below 90~MHz. We present the brightness temperature spectrum, and size of the Sun in the frequency range of 20-80~MHz.  We report on dark coronal regions with low brightness temperature that persist with frequency. The brightness temperature spectrum of the quiet Sun is discussed and compared with the bremsstrahlung emission of  a coronal model and previous quiet Sun observations.
\end{abstract}



\section{Introduction}
%[Quiet Sun in general Previous observation]
{Radio emission of the quiet Sun has been widely studied in the last few decades, pioneered by several observational studies in the microwave band.} \cite{furst1979radius} measured the radius of the Sun at centimeter wavelengths. \cite{zirin1991microwave} obtained the brightness temperature spectrum of the Sun-center with the 27-meter Owens Valley frequency-agile interferometer in the frequency range of 1.4 and 18~GHz during solar minimum.
\cite{thejappa1992unusually} found unusually low brightness temperatures ($<6\times10^4$~K) with the Clark Lake Radioheliograph in the meter-wave band.
 {The size and brightness temperature in the microwave band is found to be well described by a model considering thermal bremsstrahlung emission \citep{selhorst2005solar}. }

 {In lower frequency, the observation deviates from the model.}
\cite{subramanian2004brightness} studied the relationship between the brightness temperature and the size of the Sun at 34.5~MHz, finding that the brightness of the quiet Sun during solar a minimum (May–June 1987) ranges from $1.0 \times 10^5$~K to $4.5 \times 10^5$~K. \cite{ramesh2006equatorial} observed the equator of the quiet Sun at 51 and 77~MHz, and obtained average brightness temperatures of $3.85\times10^5$ K and $5.44\times10^5$~K, respectively. They estimated the solar angular size at the equator in the East-West [E-W] direction to be 43$^\prime$ and 38$^\prime$ at 51 and 77~MHz respectively.
\cite{mercier2015electron} imaged the quiet Sun between 150 and 450~MHz with Nan\c{c}ay Radioheliograph. They obtained the brightness temperatures and profiles in E-W and North-South [N-S] directions, and found that the electron kinetic temperature derived from radio imaging is significantly lower than the scale height temperature. \cite{vocks2018lofar} inferred a scale height temperature of $2.2\times10^6$ K with  multi-frequency interferometry imaging observation of the quiet Sun using the \textit{LOw Frequency ARray} (LOFAR) telescope's Low Band Array (LBA). \cite{melnik2018interferometric} performed interferometric imaging of the quiet Sun with the Ukrainian T-shaped Radio telescope of the second modification (UTR-2). They found brightness temperatures of $5.1 \times 10^5$~K and $ 5.7 \times 10^5 $~K at 20 and 25~MHz, respectively. Using visibility fitting, they estimated the solar size in the E-W and N-S directions to be 55$^\prime$ and 49$^\prime$ at 20~MHz and 50$^\prime$ and 42$^\prime$ at 25~MHz, respectively. 
{The Murchison Widefield Array (MWA) has been used to make comprehensive imaging observations of the quiet Sun \citep{mccauley2019low}. \citet{rahman2019relative} studied the properties}
 { of coronal holes in the frequency range of 80-240~MHz, which indicates that, compared to non-coronal hole quiet regions, coronal holes are relatively dark at high frequencies and become brighter than the background quiet Sun at the low frequency ($<145$~MHz) regime.}
\citet{ryan2021lofar} studied the spatial structure of the quiet Sun and active regions in the frequency range of 120-180~MHz with core stations of the LOFAR High Band Array (HBA). They achieved spatial resolution of 0.6 arcmin using the lunar de-occultation method; they measured the brightness temperature to be approximately $10^6$~K.
As above mentioned, most prior studies of brightness temperature spectra with spatially resolved observations of the quiet Sun have been carried out in the microwave band. Detailed, comprehensive studies in lower frequencies (meter-wave and decameter-wave) band have so far not been done, due to instrumental limitations.

% interpretation of quiet Sun emission
The quiet Sun radio emission is interpreted as thermal bremsstrahlung emission in previous studies, following the observation of \citet{zirin1991microwave}. \cite{selhorst2005solar} numerically derived the brightness temperature spectrum in the microwave band with modeled electron and ion density, and a temperature profile with altitude. Their model combines the optical observations of a photosphere model \citep{fontenla1993energy} and 0.7 times of the \cite{gabriel1992solar} transition region and corona model. The result fits the brightness temperature measurements well in the frequency range of 1.5-300~GHz. 
In lower frequency band, the size of the Sun increases with the wavelength of observation, and the observed brightness temperature is lower than that modeled \citep{mercier2015electron}, this is due to wave propagation effects, i.e. refraction and scattering. \citet{thejappa2008effects} studied the brightness temperature and size of the Sun with a ray tracing simulation. The result shows that, for quiet Sun in 50~MHz without and with scattering effects, the full width half maximum size of imaging Sun increases from 48$^\prime$ to 53$^\prime$, while the center brightness temperature is reduced by almost 61\% due to scattering with a Kolmogorov spectrum of density fluctuations and a fluctuation level $\epsilon=0.1$.
Comparing parametric simulation studies with quiet Sun observations could help constrain the plasma background properties (e.g. the fluctuation level). Thus, considering that wave propagation is frequency-dependent, further simulation-observation comparison studies require high quality and high frequency granularity observation of the quiet Sun.

% short description 
In this paper, we report the observation of the quiet Sun in the frequency range of 20-80~MHz. The results include the spatial resolved  imaging features, visual size of the Sun, and the brightness temperature spectrum. The low frequency radio imaging features are compared with EUV imaging observation of SDO/AIA \citep{SDOAIA}, the brightness temperature spectrum is discussed with a model of thermal emission.

\section{Observation and data reduction}

% instrument
We used the LOFAR-LBA outer antennas in Core+Remote observation mode for the quiet Sun observation.
LOFAR is an advanced radio antenna array \citep{van2013lofar}. It has two kinds of array observing in two frequency bands, the LBA in the frequency range of 10\,--\,90~MHz and HBA in the frequency range of 110\,--\,250~MHz. LOFAR has 52 stations, 38 of them are located in the North-East of the Netherlands, and 14 international stations are located in Germany, Poland, France, Sweden and UK. It is capable of a variety of processing operations including correlation for standard interferometric imaging, the tied-array beam-forming, and the real-time triggering of incoming station data-streams. For intense transient radio bursts the time-resolution of dynamic spectrum can reach 1/96 s, for weak and faint sources, a long time integral is applied to increase signal-to-noise ratio and obtaining a more densely distributed UV-baseline coverage.

% data acquisition
In the interferometric observation, 24 core stations and 9 remote stations were included, forming 528 station baselines, from which the longest baseline is 48~km. The digital correlator is capable of forming multiple station beams at the same time \citep{broekema2018cobalt}, enabling the simultaneous observation of multiple object with large separation angle. During the observation of this work, the primary beam is pointing at the Sun (observation object) and the secondary beam is pointing at Cassiopeia A (Cas-A, calibrator). 
We analyzed the observations on 2021-08-07 and 2021-08-14 near the local noon at the location of the LOFAR core, starting from 11:32:00.0 UT to 14:00:02.3 UT, considering the solar radio observations of decameter-wave band have better quality at noon due to the larger elevation angle of observation \citep{zhang2018type}. The rotation period of the Sun is about 25 days, so these two observations are separated by 1/4 rotation period. Thus the near limb region in the former dataset will appear as near central-disk region in the latter dataset.

% preproc
With simultaneous observation of the Sun and the Cas-A calibrator source, we perform the calibration for the visibility of each baseline using the Default Pre-Processing Pipeline (DPPP) \citep{van2018dppp}. The solution of phase and amplitude is obtained by comparing the observation of calibrator with a model of Cas-A : `\texttt{CasA\_4\_patch}' \footnote{The sky-models at ultra-low radio frequencies \citep{de2020cassiopeia} \\ \href{https://github.com/lofar-astron/prefactor/blob/master/skymodels/Ateam_LBA_CC.skymodel}{\texttt{https://github.com/lofar-astron/prefactor/blob/master/skymodels/Ateam\_LBA\_CC.skymodel}}}. Then, the solution is applied to the observations of the Sun. In this process, the difference of gain response between Sun and Cas-A observations is accounted, and thus we can obtain the absolute flux of the solar observation. Considering the static nature of the quiet Sun, we use AOFlagger\footnote{AOFlagger \citep{offringa2012morphological} \href{https://gitlab.com/aroffringa/aoflagger}{\texttt{https://gitlab.com/aroffringa/aoflagger}}} \citep{offringa2012morphological} to flag out the transient varying data frames and radio frequency interference (RFI); this flagging step can significantly improve the quality of imaging for static radio sources.

% imaging
Next, the calibrated visibility in the form of measurement-sets is input into WSClean\footnote{WSClean \citep{offringa2014wsclean} \href{https://gitlab.com/aroffringa/wsclean}{\texttt{https://gitlab.com/aroffringa/wsclean}}} \citep{offringa2014wsclean} for the procedures of UV-space sampling, Fourier transformation, and deconvolution.
In order to reveal the weak and faint structure in the imaging, we applied the full 2.45 hours' time-integration duration and 0.195MHz frequency averaging to increase signal to noise ratio,  { as well as obtain a denser distributed UV coverage}. This integration can be applied because we are studying the quiet Sun emission, which is varying on significantly longer time scales than solar radio transients, e.g. radio bursts associated with the eruptive processes, such as flares and coronal mass ejections (CMEs). The multi-scale option of WSClean is activated for the imaging of spatially extended structures.
The observation duration is 8882.33 seconds for each of these two datasets. The starting 55 seconds and ending 15 seconds of observation are not included in the time-integration. The resulting image is in astronomical coordinates and units. We converted the coordinates into Helioprojective coordinates and the flux density into brightness temperature using the LOFAR-Sun-tools\footnote{LOFAR-Sun-tools \citep{zhang2020interferometric} \href{https://git.astron.nl/ssw-ksp/lofar-sun-tools}{\texttt{https://github.com/peijin94/LOFAR-Sun-tools}}} library.

\section{Results}
\begin{figure}
	\centering
	\includegraphics[height=0.4\linewidth]{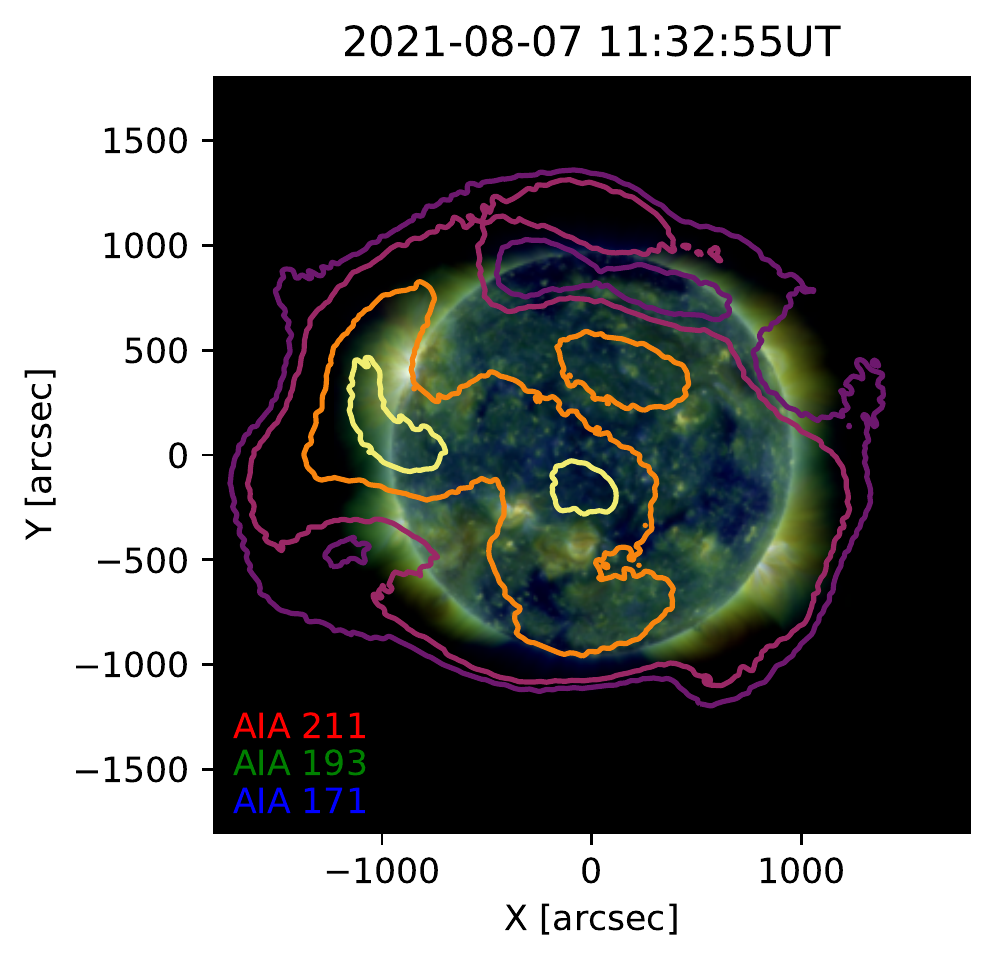}
    \includegraphics[height=0.4\linewidth]{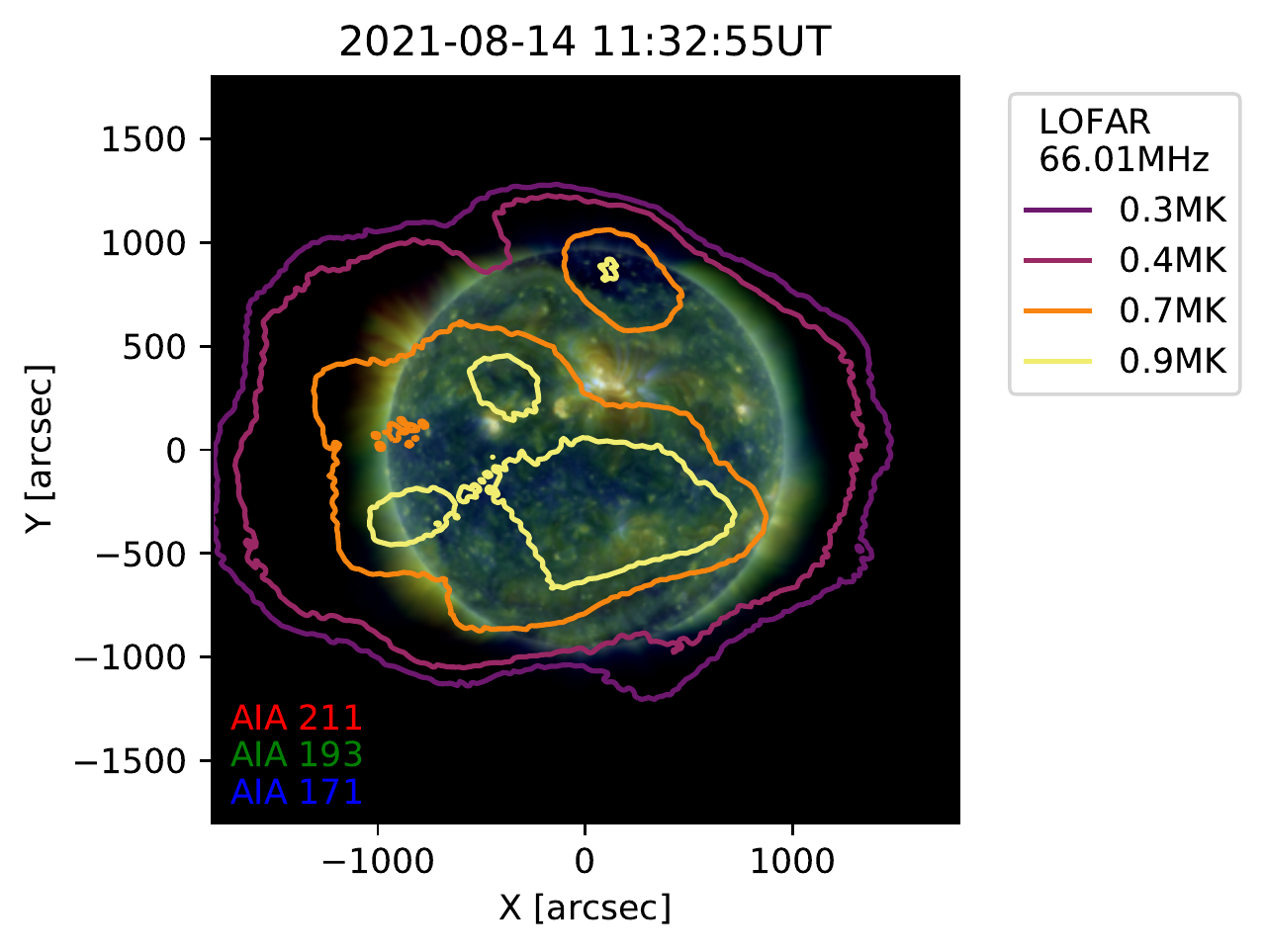}
	\caption{The contour of brightness temperature distribution from LOFAR-LBA at 66.01~MHz overlapped to the image of the composite tri-color image of EUV image from SDO/AIA, composed of the wavelength of 211\AA (red) , 193\AA (green) , 171\AA (blue).}
	\label{fig:sdo}
	\end{figure}

% Qualitative description
From the observations, we obtained one image on each day in each subband.
% Overlap
Figure \ref{fig:sdo} shows the contour of radio brightness temperature distribution of the 66~MHz observations, overlapped onto an EUV image of SDO/AIA \citep{SDOAIA}.
% Active region
The quiet Sun emission on the 2021-08-07 has a contribution from the near-east limb active regions NOAA AR 2853 and NOAA AR 2852, situated at that moment close to the central meridian. During the second considered day, i.e. 2021-08-14, the quiet Sun emission is dominated by the contribution from the NOAA AR 2853 and with somewhat smaller contribution of the NOAA A 2855 (both of these active regions are situated close to the central meridian).
By comparing the left and right panels of in Figure \ref{fig:sdo}, one can see that the active region is brighter on the limb than in the center.
Both the northern polar coronal hole, and equatorial coronal hole are bright on both these days.

\begin{figure}
	\centering
	\includegraphics[width=0.99\linewidth]{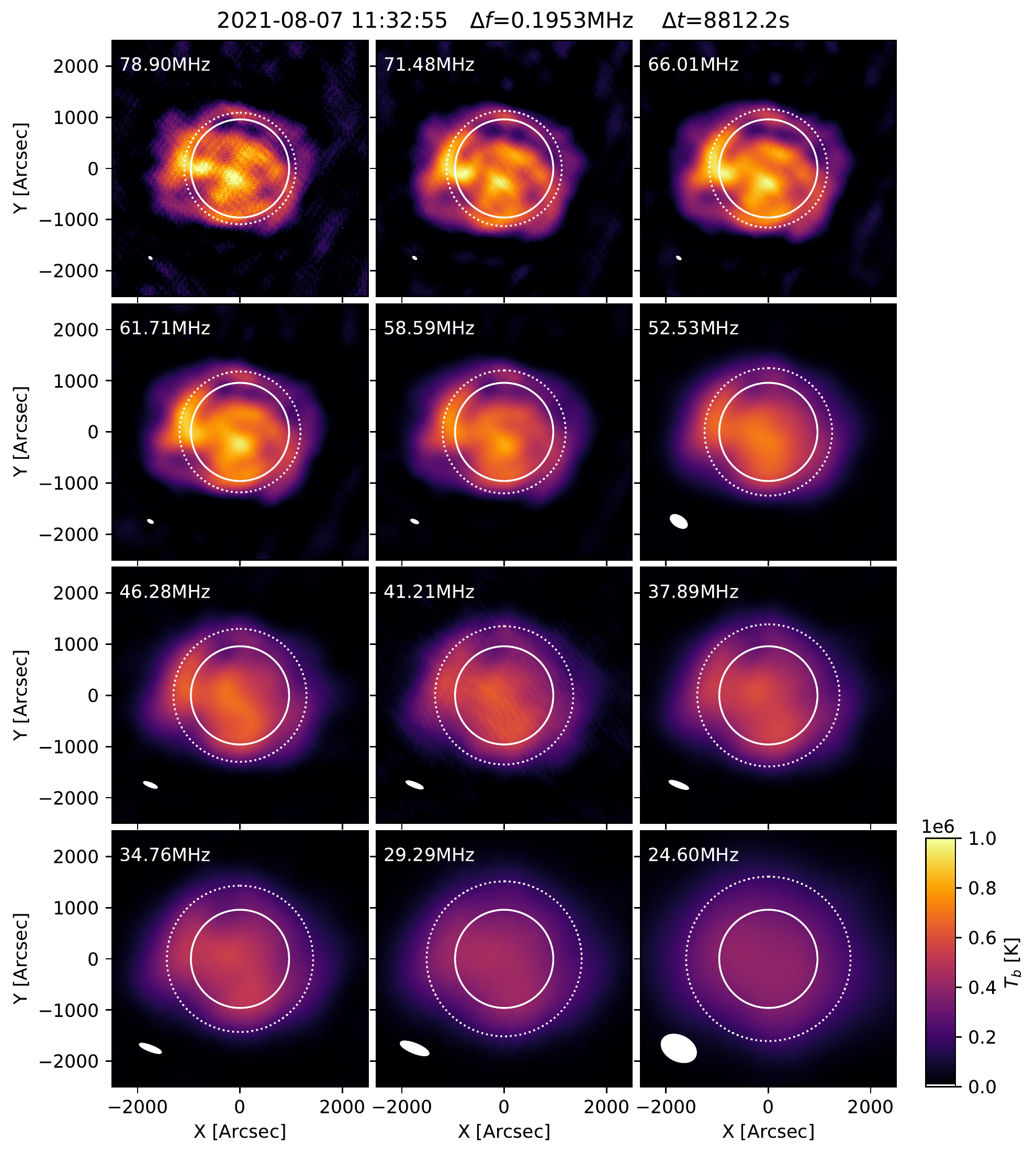}
	\caption{The interferometry imaging of the quiet Sun on 2021-08-07. The beam shape is marked as a white patch in the lower left corner of each panel; solid white lines represents the edge of the optical solar disk; dotted white line represents the local frequency plasma height ($f=f_{pe}(h)$) according to \citep[equator]{saito1977study} electron density model.}
	\label{fig:main:1}
	\end{figure}

\begin{figure}
	\centering
	\includegraphics[width=0.99\linewidth]{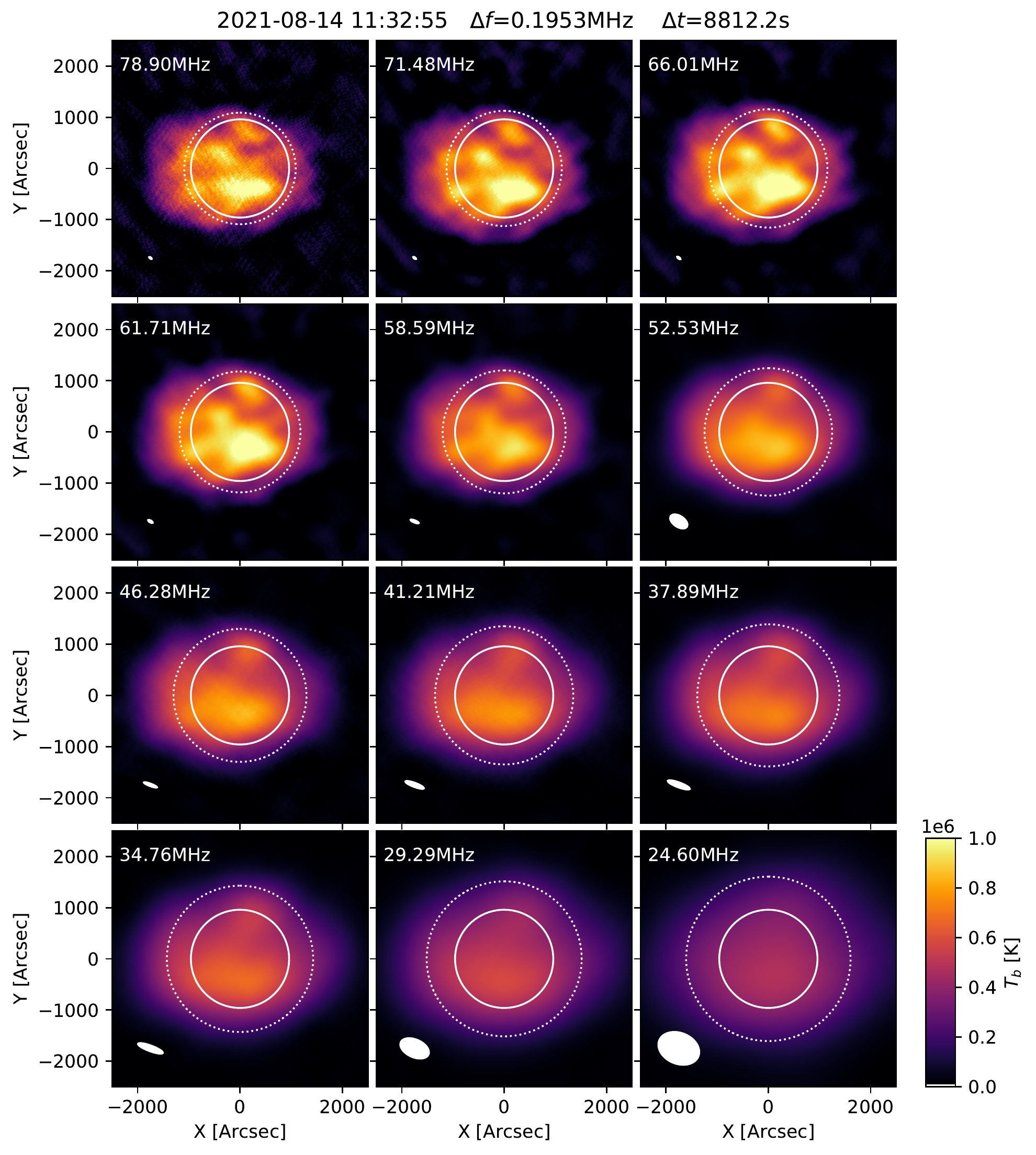}
	\caption{The interferometry imaging of the quiet Sun on 2021-08-14. The beam shape is marked at white patch in the lower left corner of each panel; solid white line represents the edge of the optical solar disk; dotted white line represents the local frequency plasma height ($f=f_{pe}(h)$) according to \citep[equator]{saito1977study} electron density model.}
	\label{fig:main:2}
	\end{figure}

% main figures
Figures \ref{fig:main:1} and \ref{fig:main:2} show the radio images of the Sun at 12 selected subbands on 2021-08-07 and 2021-08-14. As a reference, the radius of local plasma frequency ($f=f_{pe}(h)$) according to the density model of {Saito77 \citep[shown in Eq. \ref{eq:saito}]{saito1977study}} is marked as white dotted circles. As shown in Figures \ref{fig:main:1} and \ref{fig:main:2}, the quiet Sun emission extends out of the regime of local plasma frequency in all directions, especially in the horizontal [E-W] direction. 
% spatial scales
The images in Figures \ref{fig:main:1} and \ref{fig:main:2} show that there are rich spatial details in arc-minute scale in the frequency range of above 60~MHz, from which some dark regions with low brightness temperature are visible.
The spatial details gradually fade out with frequency decreasing to the frequency range of 30~MHz, where only degree-scale spatial structures can be resolved.

\subsection{Brightness temperature spectrum}

In the following sub-sections, we will provide quantitative descriptions of the brightness temperature spectrum of the full Sun, including the spectra of different regions, as well as the brightness temperature slice profiles along the E-W and N-S diameters of the quiet Sun, and its variation with frequency.

\begin{figure}[h]
	\centering
	\includegraphics[width=0.75\linewidth]{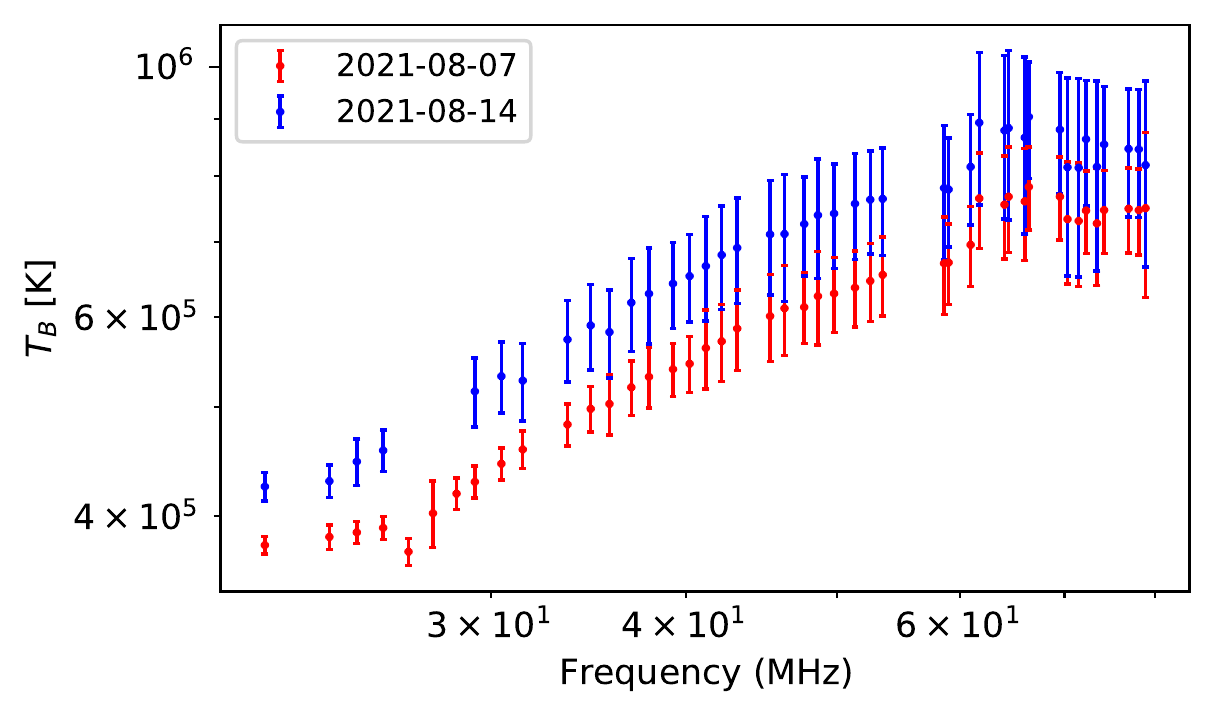}
	\caption{Brightness temperature spectrum of the center region on 2021-08-07 and 2021-08-14. The error bars represent the standard deviation of the values within $r<0.5R_{sun}$.}
	\label{fig:tbspec2day}
	\end{figure}

We measure the quiet Sun disk center brightness temperature spectrum by calculating the average value in the region of $r<0.5 R_{sun}$ from the disk center for each subband, where $R_{sun}$ is the solar optical radius. We obtained 45 subbands on 2021-08-07 and 42 subbands on 2021-08-14. The frequency and average brightness temperature are shown in Table  \ref{long} and \ref{long2} and Figure \ref{fig:tbspec2day}. The error range is given by the standard deviation of brightness temperature in the region. From Figure \ref{fig:tbspec2day}, the brightness temperature of the quiet Sun disk center in the frequency range of 20-80~MHz  is higher on 2021-08-14 than on 2021-08-07. Considering the brightness temperature distribution in Figures \ref{fig:main:1} and \ref{fig:main:2}, the difference in the spectra between 2021-08-07 and 2021-08-14 is mainly due to the bright coronal hole near the equator on 2021-08-14.

\begin{figure}[h]
	\centering
	\includegraphics[width=0.9\linewidth]{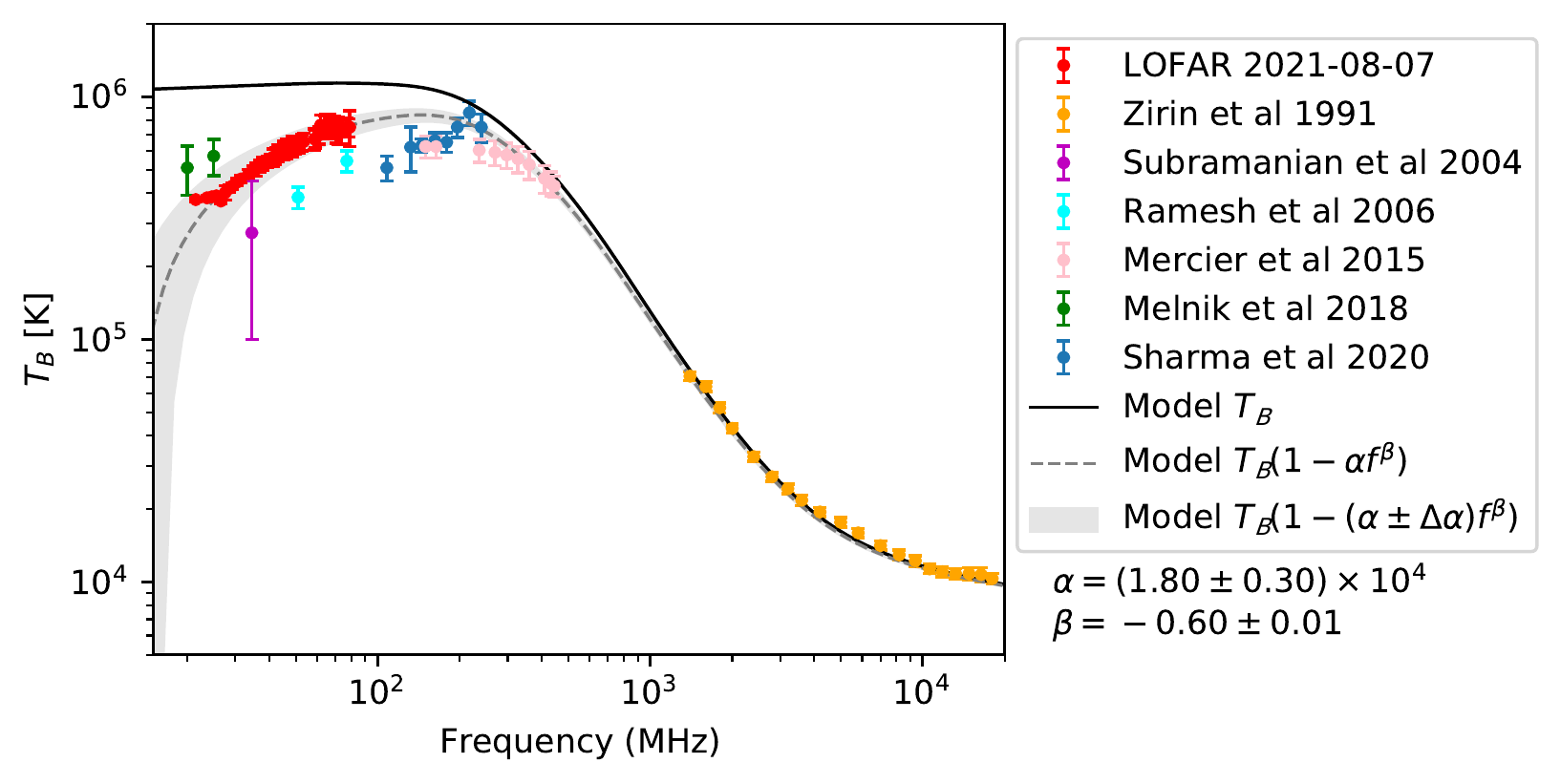}
	\caption{The brightness temperature spectrum of LOFAR on 2021-08-07 (red), and previous works: \cite{zirin1991microwave}, \cite{subramanian2004brightness},
	\cite{ramesh2006equatorial},\cite{mercier2015electron},
	\cite{melnik2018interferometric}, \cite{sharma2020propagation}. 
	The brightness temperature in this work is obtained by averaging the brightness of $r<0.5R_{Sun}$ region in the imaging of each frequency. The black continuous line is the modeled brightness temperature of bremsstrahlung emission \citep{selhorst2005solar}; the gray dashed line is the mean fitted flux attenuation ratio due to propagation effects; the gray shaded area shows the effect of the variation in the fit parameters. The data points of LOFAR observation in this figure are available in Table \ref{long}.}
	\label{fig:tbspec}
	\end{figure}

To compare the observed brightness temperature spectrum with the model, we extended the model of \cite{selhorst2005solar,selhorst2019solar} into lower frequencies (The details are described in Appendix A).
The modeled spectrum and observation results are shown in Figure \ref{fig:tbspec}, including results of 2021-08-07 in this work and previous measurements of the quiet Sun brightness temperature.
We used the data of 2021-08-07 instead of 2021-08-14 - this is done to avoid the bright region of a large coronal hole near the disk center on 2021-08-14. We intend to use non-coronal-hole quiet Sun emission in the comparison.
The modeled spectrum (shown as solid black line in Figure \ref{fig:tbspec}), shows a brightness temperature approaching $10^6$ K in the decameter-wave band.

Comparing the observation with the model, we find consistency with the model spectrum of the microwave band ($>1$~GHz) \citep{zirin1991microwave}, while in the meter-wave and decameter-wave band ($<400$~MHz), the observed brightness temperature is significantly lower than the model, and the difference between the model and observation increases with the decrease of frequency.

As the propagation effects (i.e. the refraction and scattering) are frequency-dependent, we use a power-law term to represent and fit the attenuation of brightness temperature due to propagation effects, expressed as:

\begin{equation}
    T_B'(f)  = T_B(f) (1-\alpha f^\beta)\quad,
    \label{eq:tbfit}
\end{equation}
where $T_B(f)$ is the modeled brightness temperature spectrum (solid black line in Figure \ref{fig:tbspec}), and $f$ is the frequency in unit of Hz. By fitting the LOFAR observation data on 2021-08-07 to Equation \ref{eq:tbfit}, we obtain $\alpha=(1.80\pm0.30)\times 10^4$ and $\beta=-0.60\pm0.01$, shown as the gray dashed line in Figure \ref{fig:tbspec}. The fitting result is well consistent with the previous measurements of quiet Sun imaging in literature (as shown in Figure \ref{fig:tbspec}).

\begin{figure}[h]
	\centering
	\includegraphics[width=0.85\linewidth]{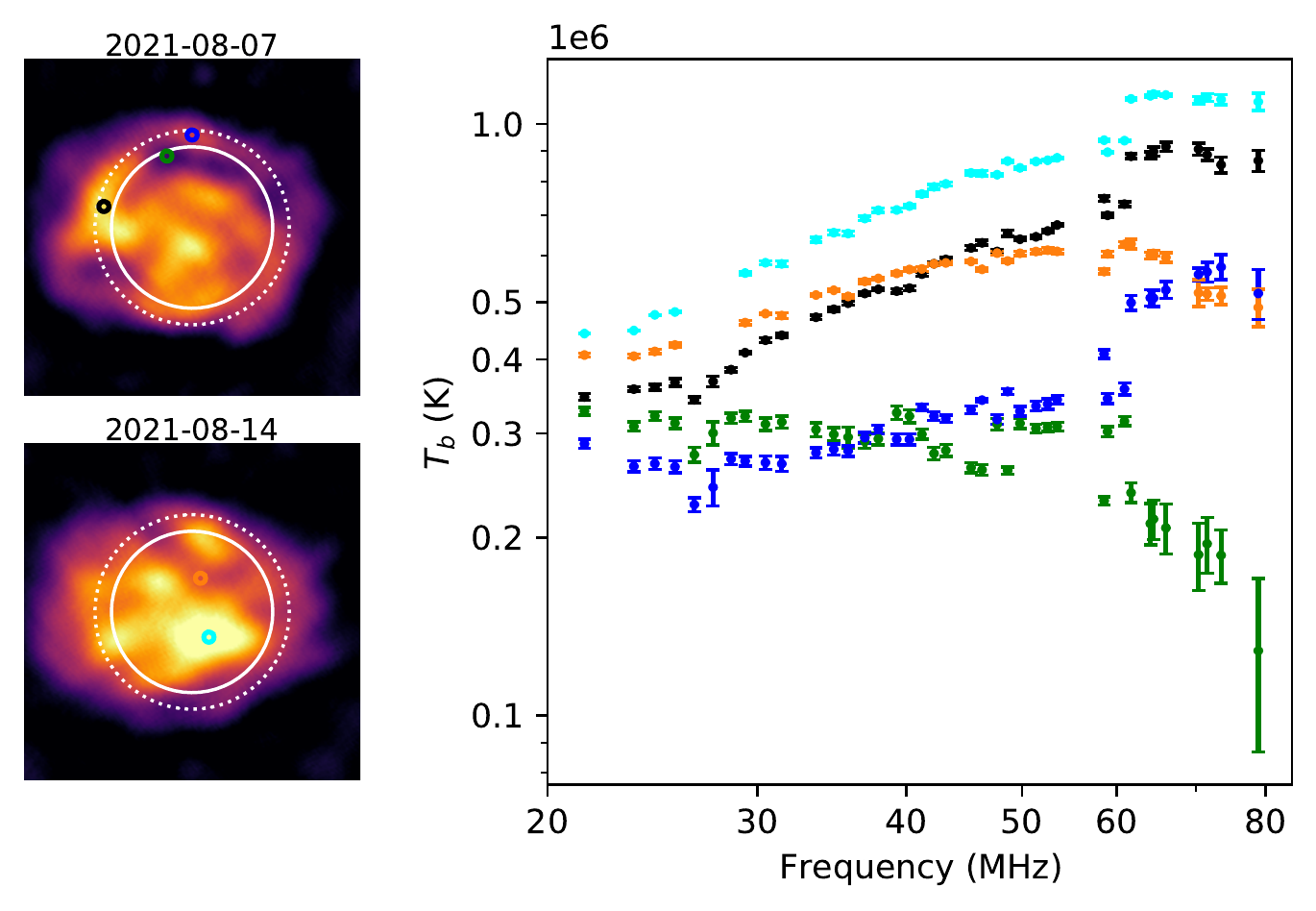}
	\caption{Brightness temperature spectrum of given points on 2021-08-07 and 2021-08-14. Colored circles on the left two panels marks the position of the coordinate points, the size of the circle represents the regime of averaging. The brightness temperature spectrum is shown in the right panel with corresponding color.}
	\label{fig:clips}
\end{figure}

With imaging of multi-frequency subbands, we can obtain the  brightness temperature spectrum of given coordinate points. Figure \ref{fig:clips} shows the brightness temperature of 5 positions on 2021-08-07 and 2021-08-14. The five positions in Fig. \ref{fig:clips} include five  {representative} features: the green point represents the low brightness temperature dark region, the blue represents the bright patch above the Northern-polar coronal hole, black represents the active region on the limb, the orange represents the active region near the disk, and the cyan represents the bright equatorial coronal hole.
% description

The spectra for these points show that the brightest region observed during these two days is the equatorial coronal hole, reaching $10^6$ K in the frequency range of 60-80~MHz. Comparing the same active region at the eastern limb (black) and the point near the disk center (orange), we find that the spectrum of the limb active region is steeper than the center active region: the limb active region has a higher brightness temperature than the disk center active region in the frequency range of 40-80~MHz, but lower brightness temperature in the lower frequency range (20-40~MHz). Comparing the points near the coronal hole on 2021-08-07 (green and blue) in Fig. \ref{fig:clips}, we find that, although these two positions are only 6 arcmin apart, there is a large difference in the brightness temperature spectrum. The bright patch has a positive slope spectrum, while the dark region has a negative slope spectrum. The largest deviation of brightness temperature of these two points happens at 78~MHz (the highest frequency subband in the dataset), where the dark region is $0.13\times10^6$ K and the bright patch is $0.50\times10^6$ K. The brightness temperature of all these five points converges to 0.25-0.45$\times10^6$~K at the low frequency end (21.5~MHz) of the observations.  {This convergence 
of brightness temperature represents a smooth spatial distribution in low frequency  channels ($<$30~MHz), which we assume is {mainly} due to scattering of radio waves.}

\subsection{Size of the Sun in Low Frequency}

	\begin{figure}[h]
	\centering
	\includegraphics[height=0.66\linewidth]{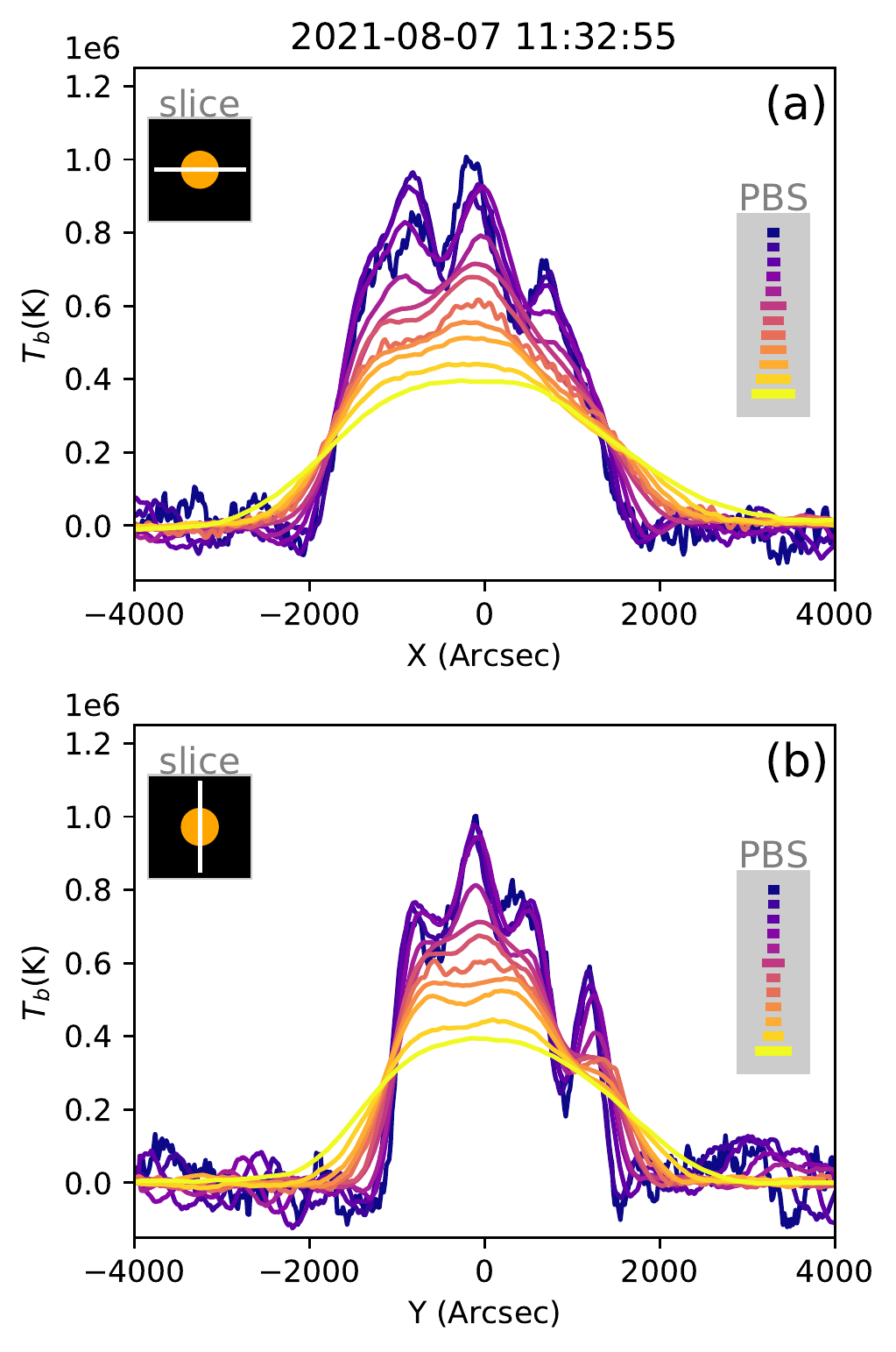}
    \includegraphics[height=0.66\linewidth]{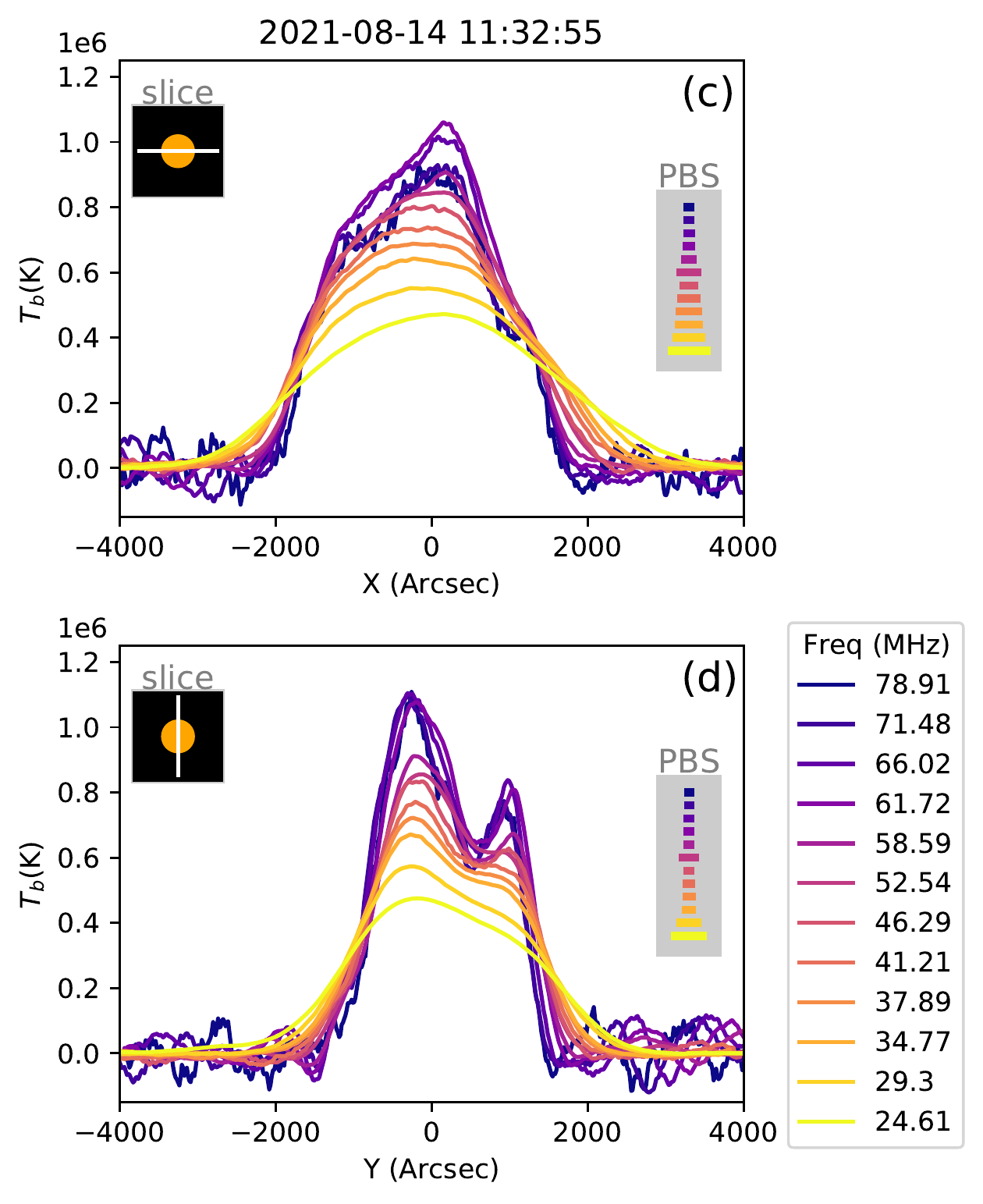}
	\caption{Slices of brightness temperature distribution along the center line of longitude and latitude. Projected beam size (PBS) represents the beam-size projected to the slice line.}
	\label{fig:slice}
	\end{figure}	

{The size of the Sun in low-frequency images is not well-defined due to the complex $T_B$ distribution. Different measuring methods give different definitions to the delimitation, which leads to different size measurement results.
There are a variety of methods used in previous studies of the quiet Sun. A half-power method was used by \citet{aubier1971observations}. The signal is first corrected by the formula $\theta_S=\sqrt{\theta_0^2-\theta_b^2}$, where $\theta_0$ is the measured scan, $\theta_b$ is the beam-width, $\theta_S$ is the corrected source size. The size measured by this method can ignore extended solar radio emission below the half-power level.
The Gaussian fitting method assumes the radio source to have a Gaussian-like flux distribution, and fits it with a Gaussian function in UV-visibility  \citet{melnik2018interferometric} or image space \citet{ramesh2006equatorial}. The size of the radio Sun is thus determined by the full width at half-maximum of the fitted Gaussian.
Finally, the threshold-above-background-level method was used by \cite{sharma2020propagation}. The threshold of this method delimits the solar radio emission and background, thus the measurement results of this method can represent the size of the region with solar radio emission detected.}
%, while for this method it is hard to make corrections for the influence of non-zero beam-size

\begin{figure}[h]
	\centering
	\includegraphics[width=0.5\linewidth]{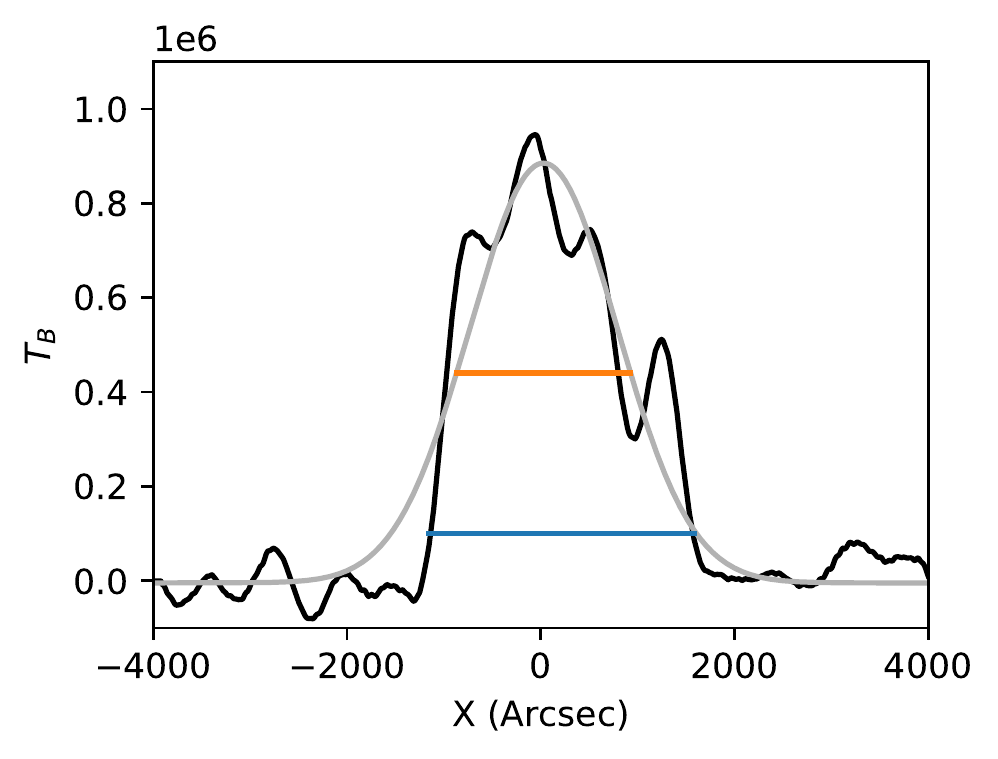}
	\caption{ $T_b$ slice in NS direction of 61.72MHz (black solid line) on 2021-08-07, and width measurement results, gray curve represents the Gaussian fit, orange-line marks the width measured by FWHM of Gaussian-fit, blue-line marks the width measured by threshold of 0.1~MK.}
	\label{fig:radius_demo}
\end{figure}

{In this work,} the effective radius is measured from the brightness temperature slice profile at the center longitude and latitude lines. As an example, the slice of 12 subbands is shown in Figure \ref{fig:slice}. The frequency subband is the same as in Fig. \ref{fig:main:1}
 and \ref{fig:main:2}. 
From Fig. \ref{fig:slice}, it is apparent that the profile is wider  {in the equatorward direction than that of the central meridian}, and the width increases with decreasing frequency. The beam-size in the central meridian line is smaller than that of the equator line. As the beam-size is significantly smaller than the profile width (solar diameter), we can obtain relatively precise measurements on the solar radius from the observed profiles.
{
%In this work, 
We use the threshold-above-background-level method to present the size of the region where solar emission is detected, and we also provide the results from Gaussian fit as comparison. The effective radius ($R_{0.1\rm MK}$) is measured as the half-width of the regime above the 0.1~MK threshold, whereas the effective radius of Gaussian-fit ($R_{\rm fit}$) is measured as the half-width of the fitted distribution. The results are shown in Table \ref{long} and \ref{long2} for observation on 2021-08-07 and 2021-08-14  respectively. An example of these two method is shown in Figure \ref{fig:radius_demo}, as we can see in this case, the size measured by Gaussian fit is smaller than the 0.1~MK-threshold methods.}

\begin{figure}[h]
	\centering
	\includegraphics[width=0.8\linewidth]{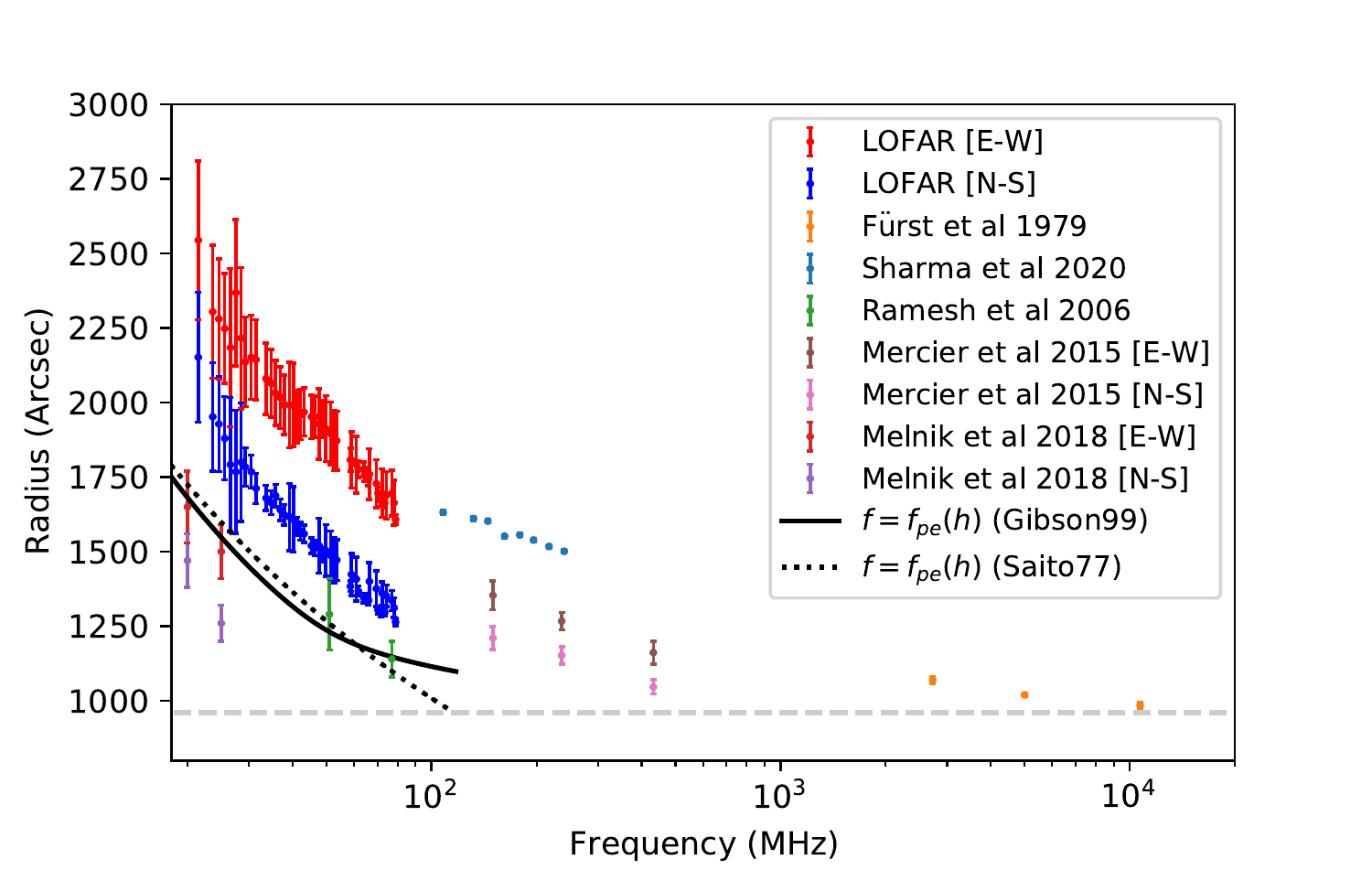}
	\caption{The effective size of the Sun measured  {0.1~MK} across the observed frequency band on 2021-08-07,  {errorbars represent the projected beam size}. Data points of LOFAR observation in this figure (red for E-W, blue for N-S) are available in Table \ref{long}.  {The black solid and dashed line represents the local plasma radius from the density model of \cite{saito1977study,gibson1999solar}}, and the gray dashed line represents the optical radius. Results from other studies are given for comparison.}
	\label{fig:radius}
\end{figure}

We obtained the effective radius of quiet Sun in 45 subbands on 2021-08-07 and 42 subbands on 2021-08-14. The values are presented in Table \ref{long} and Table \ref{long2}, and shown in Figure \ref{fig:radius}. The width in the horizontal [E-W] direction is significantly larger than the width in the vertical [N-S] direction. This could be related to the higher plasma density of the  coronal streamers in the low-middle latitude with respect to the lower plasma density of the coronal holes in the polar region. The effective radius observed by LOFAR follows the same trend as the local plasma radius ($f=f_{pe}(h)$). 
 {In Figure \ref{fig:radius}, we put the local plasma radius of two density models as reference, the equator model \citep{saito1977study} expressed as: 
}\begin{equation}
    N_e(r) = 1.36\times10^6  r^{-2.14} + 1.68\times10^8  r^{-6.13}\quad[\rm cm^{-3}]\quad,
    \label{eq:saito}
\end{equation}
 {and the streamer model \citep{gibson1999solar} expressed as:}
\begin{equation}
    N_e(r) = (77.1  r^{-31.4} + 0.954  r^{-8.30}+ 0.550  r^{-4.63})\times10^8 \quad[\rm cm^{-3}]\quad.
\end{equation}
From Figure \ref{fig:radius}, we can see that the effective radii in N-S direction is  closer to local plasma radius. 
The solar size in N-S direction are slightly larger than the result from MWA observations of quiet Sun equator by \cite{ramesh2006equatorial}. Both E-W and N-S radii are significantly larger than the result of \cite{melnik2018interferometric}. The trend of radius-frequency to high frequency is consistent with \cite{mercier2015electron} and \cite{sharma2020propagation}.

\section{Discussion}

% Center Brightness temperature 
The brightness temperature spectrum obtained in this work is consistent with previous works in the frequency range of 20-80~MHz. We must mention that the results from \citet{ramesh2006equatorial} were observed in 1995 during solar minimum, while the results from \citet{melnik2018interferometric} were observed in 2014 during solar maximum. In this work, the observation time is 2021-August, during the rising phase of solar activity early in the new cycle.  {The solar-cycle variation in brightness temperature} requires further long-term calibrated observations.
The observed center brightness temperatures of low frequency ($<100$~MHz) in this and previous work are all below the model predicted brightness temperature value of $10^6$ K. 
In the frequency range of 150-400~MHz \citep{mercier2015electron}, the observed brightness temperature ($10^5$ K) is below the modeled brightness temperature (black solid line in Figure \ref{fig:tbspec}) and the scale height temperature ($2.2\times10^6$ K). 
% fill gap of 1G
The observed brightness temperature of the Sun has a gap near 1~GHz. Future work of calibrated observations to measure the brightness temperature and size in this frequency range \citep[]{yan2021mingantu} could help to complete the spectrum. 
% compare model

We assume the attenuation of the flux is due to the propagation effect. The model (shown in Appendix A) considers the free-free emission and absorption in a straight line of sight (LOS), while in meter-wave and decameter-wave band, refraction and scattering can significantly  {deviate the wave propagation path from the straight line, and thus influence} the imaging results \citep{tan2015study,kontar2019anisotropic,zhang2021parametric}. The Monte Carlo simulation  {of radio wave propagation ray tracing} \citep{thejappa2008effects} with  {the plasma background density fluctuation is described by two parameter combinations : (1) a spectral index of 11/3 representing Kolmogorov spectrum with fluctuation level $\epsilon=0.1$; (2) a spectral index of 3, representing a flat spectrum with fluctuation level $\epsilon=0.02$. The resulting center brightness temperature is $0.35$~MK and $0.4$~MK respectively for 50~MHz, which is lower than the LOFAR observation obtained in this work of $0.63$~MK for 49.8~MHz.}

% spectrum of points
From the spectrum of several coordinate points (as shown in Figure \ref{fig:clips}), one can see the  {large variation} of the brightness temperature -- at the frequency range of 60-80~MHz, it ranges from 0.15 to 1~MK.  {The equator coronal hole region (cyan in Figure \ref{fig:clips}) is significantly brighter than the background in all frequency channels. This is consistent with previous MWA observations \citep{mccauley2019low} which confirms that for <145~MHz, coronal holes at disk center tend to be brighter than the background quiet Sun}. The interpretation of this  {large variation in} brightness temperature may require an inhomogeneous solar atmosphere temperature and density distribution, or introduction of magnetic field and cyclotron emission into the interpretation.
% active region
The spectrum of the active region on the two days (shown as black and orange in Figure \ref{fig:clips}) shows that the active region is brighter on the limb in 40-80~MHz, while the center active region is brighter in 20-40~MHz. We suggest that this occurs because in lower frequency, the scattering effect  {diffuses the emission from} the bright coronal hole to a larger region.  The scattering from the inhomogeneous plasma background could blur the imaging and increase the scale of the smallest observable structure. 
In future work, comparing  {the spatial scale of the smallest structure in simulations of wave propagation and  imaging observation} (e.g. this work and \cite{ryan2021lofar}) could help quantify the scattering parameters, such as density fluctuation amplitude and the degree of anisotropy.

\begin{figure}
    \centering
    \includegraphics[width=0.9\linewidth]{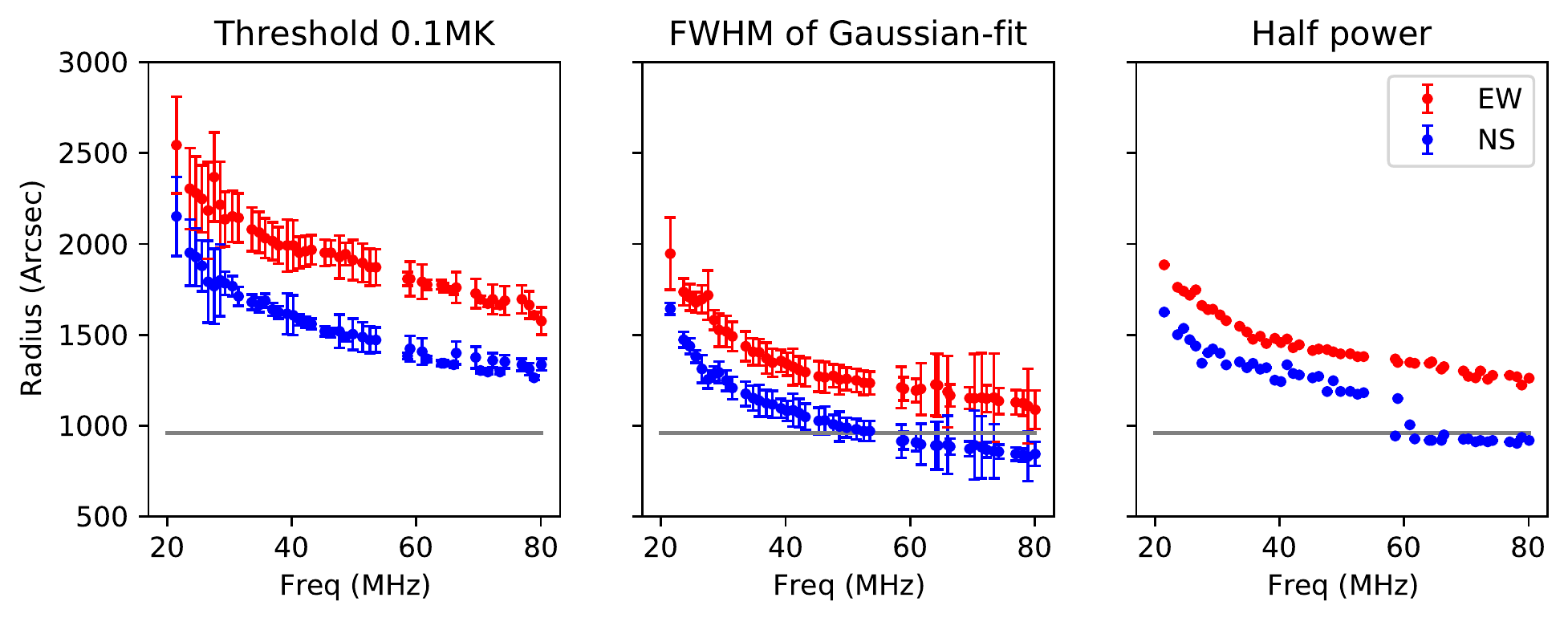}
    \caption{The effective radius measured by different methods, for the observation of 2021-08-07, the gray line represents the optical radius. }
    \label{fig:width_comparison}
\end{figure}

% Size of the quiet Sun
{The brightness distribution of the Sun at low frequencies is complex and has a vague undetermined edge, which makes the size measurement challenging. 
There are several methods used in previous studies as introduced in Section 3.2. Different methods give different definitions to the delimit and are measuring different parts of the Sun. The half-power and FWHM of Gaussian-fit method measures the size of the major emission region, the threshold-above-background methods measure the size of the region with emission detected. Figure \ref{fig:width_comparison} presents the result of three methods, as we can see that, the result with value  threshold 0.1~MK is larger than that of Gaussian-fit and half-power threshold method. The Gaussian-fit and half-power method have similar results in EW direction. The result of half-power in NS direction is not continuous in the frequency range of 50-70~MHz because when there are dark regions near the edge, the threshold of half-power may exclude some secondary peaks near the threshold. As comparison, the results from Gaussian-fit and 0.1~MK threshold methods are continuous in frequency.
}
The size of the Sun in Figure \ref{fig:radius} is measured {at 0.1~MK threshold  {(which is about the background noise level)} in} the slice profile at E-W and N-S direction. 
{We need to notify that, the radius measured with this method could be influenced by the beam-size.} 
Comparing the results in this work with previous studies,
the radius at 0.1~MK in this work has a trend which is consistent to the observation of NRH by \cite{mercier2015electron} and \cite{sharma2020propagation}. 
{The result of Gaussian-fit measurements in this work is consistent with the previous  Gaussian-fit measurements in brightness distribution \cite{ramesh2006equatorial} and the radius measurement result by fitting UV-visibility distribution to Gaussian distribution \citep{melnik2018interferometric}. This indicates the equivalence of these two methods. We should expect similar size measurement results as Gaussian fitting in image-space and UV-space are theoretically equivalent. } 
{As we can see from the above size comparison: the size from Gaussian fit is smaller than the size from 'threshold-above-background-level'. This is due to the difference of the  threshold value and curve shape (slice profile and its Gaussian-fit).}  As the quiet Sun is a complex source as shown in Fig. \ref{fig:main:1} and Fig. \ref{fig:main:2}, which can't be well described by Gaussian distribution, thus, determination of its diameter with Gaussian fitting could result in underestimation of the solar radius. 
% compare to local plasma frequency
The size of the Sun in this work is larger than the local plasma radius with \cite{saito1977study} density model, especially in E-W direction. The distance of the emission from the region of local plasma frequency is an important parameter to consider, because the propagation effect decreases with distance from the local plasma radius. The size and spatial distribution could provide a reference for future wave propagation study of quiet-Sun emission.

\section{Conclusions}

In this work, we use LOFAR-LBA to observe the quiet Sun. The major results include:
\begin{itemize}
    \item We measured the disk center brightness temperature and the radius of the quiet Sun in frequency range of 20-80~MHz on 2021-08-07 and 2021-08-14, detailed shown in Table \ref{long} and Table \ref{long2}.
    \item The brightness temperature is consistent with previous work in this frequency range, but lower than modeled values. We assume the difference is due to propagation effects. The brightness temperature on 2021-08-07 could be fitted by $ T_B'(f)  = T_B(f) (1-\alpha f^\beta) $ with $\alpha=(1.80\pm0.30)\times 10^4$ and $\beta=-0.60\pm0.01$.
    \item The radius of the quiet Sun is measured at 0.1~MK. The results are consistent with previous measurements: the radius in E-W direction is larger than that in N-S direction, and both are larger than the local plasma frequency radius.
\end{itemize}

The brightness temperature and size observations in this work provides a reference for future parametric simulation works. More comprehensive observation-targeted parametric simulation works could help to diagnose the plasma background by searching for the best match parameter combinations to reproduce the observations.

\section*{Acknowledgements}
% this work
This work is supported by the European Union's Horizon 2020 research and innovation programme under grant agreement 952439, project STELLAR (Scientific and Technological Excellence by Leveraging LOFAR Advancements in Radio Astronomy). K. Kozarev acknowledges support from the Bulgarian National Science Fund, VIHREN program, under contract KP-06-DV-8/18.12.2019. 
% Bartosz
B. Dabrowski and A. Krankowski thank the National Science Centre, Poland and the Beethoven Classic 3 funding initiative under project number 2018/31/G/ST9/01341, the Ministry of Education and Science (MES), Poland for granting funds for the Polish contribution to the International LOFAR Telescope (agreement no. 2021/WK/02)
% Peijin
The research at USTC is supported by the National Natural Science Foundation of China (41974199 and 42188101) and the B-type Strategic Priority Program of the Chinese Academy of Sciences (XDB41000000).
% lofar
The data in this work is obtained with the International LOFAR Telescope (ILT) under project code \texttt{LT16\_001}. LOFAR is the Low Frequency Array designed and constructed by ASTRON. It has observing, data processing, and data storage facilities in several countries, that are owned by various parties (each with their own funding sources), and that are collectively operated by the ILT foundation under a joint scientific policy. The ILT resources have benefited from the following recent major funding sources: CNRS-INSU, Observatoire de Paris and Universite d'Orleans, France; BMBF, MIWF-NRW, MPG, Germany; Science Foundation Ireland (SFI), Department of Business, Enterprise and Innovation (DBEI), Ireland; NWO, The Netherlands; The Science and Technology Facilities Council, UK; Ministry of Science and Higher Education, Poland."

\appendix
\section{Model of brightness temperature spectrum}
The brightness temperature spectrum of the quiet Sun can be numerically derived from a model of density and temperature of solar atmosphere considering the bremsstrahlung radiation, by numerically integrating the radiation transfer equation along line of sight \citep{selhorst2005solar,tan2015study,selhorst2019solar}:

\begin{equation}
    T_B (f) = \int_{\rm LOS} \it T \kappa_f e^{-\tau_f} \rm{d}\it s .
\label{eq:tb}
\end{equation}

\begin{figure}[h]
	\centering
	\includegraphics[height=0.63\linewidth]{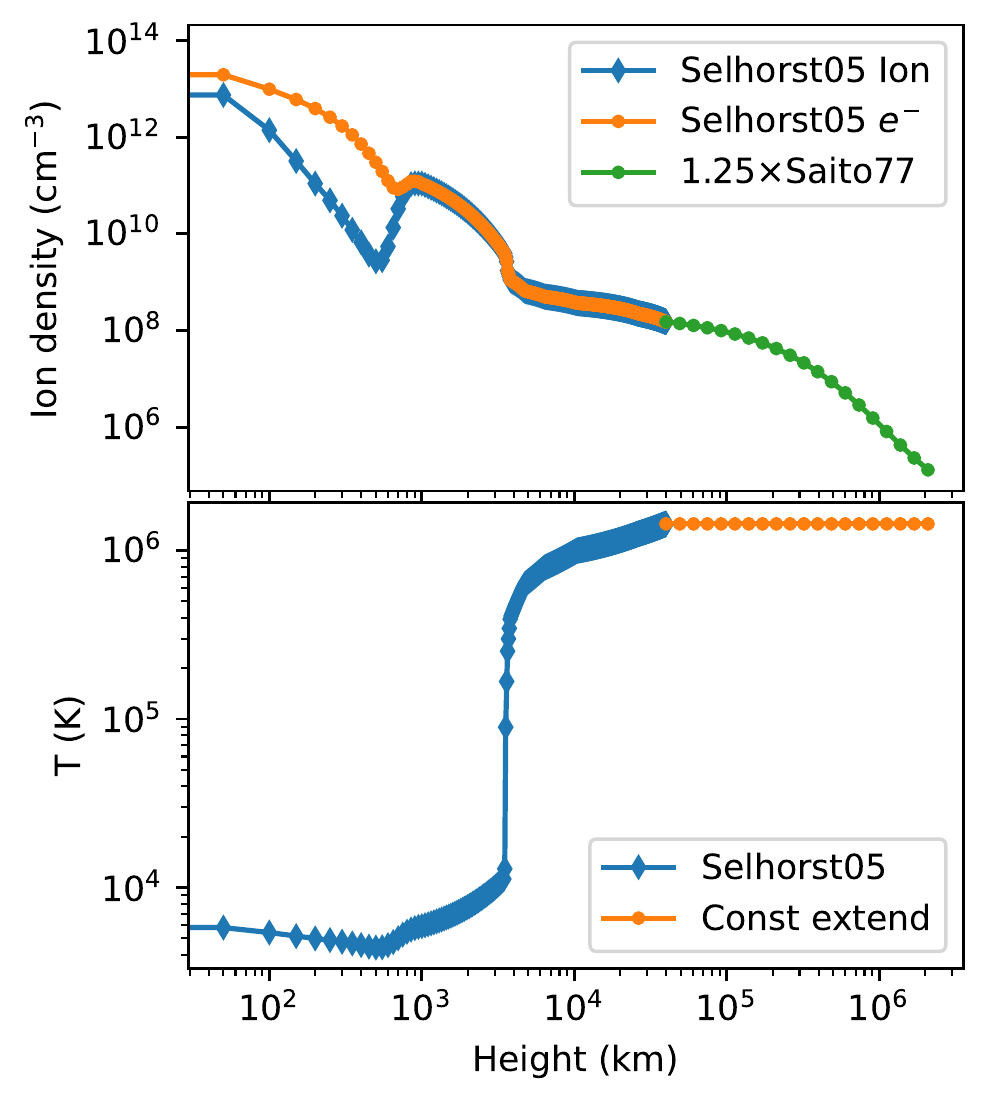}
	\caption{Variation of the plasma density and temperature  with height in the solar atmosphere, which is used for modeling the radio-wave brightness temperature spectrum.}
	\label{fig:apd:1}
	\end{figure} 
Where $T$ is the temperature of the medium, for which we use \cite{selhorst2005solar} model with constant extension for the outer corona, as there is no rich information for the temperature of the outer corona in literature, and $10^6$ K is a good estimation.
In equation \ref{eq:tb}, $\tau_f(s_0)=\int_{s_0}^{\inf} \kappa_f(s) \rm{d}\it s$ is the optical depth, $\kappa_f$ is the opacity, expressed as:

\begin{align}
    &\kappa_f = 3.7\times10^8 \it T^{-1/2}n_e n_i f^{-3}g_{ff}(1-e^{-h f/k_b T}), 
\end{align}
where, $n_e, n_i$ is the density of electrons and ions in the solar atmosphere and outer corona, for which we use the model of \cite{selhorst2005solar} below 40Mm altitude, and 1.25 times the density model of Saito77 \citep{saito1977study} for the outer corona, as shown in Figure \ref{fig:apd:1}. $g_{ff}$ is the Gaunt factor, given by \cite{van2014accurate}.

With the model described above, a brightness temperature spectrum ($T_B(f)$) can be obtained by numerically solving Equation \ref{eq:tb} for a series of frequencies, as shown in Fig. \ref{fig:tbspec}. 

\section{Brightness temperature and size in observation}

Table \ref{long} and \ref{long2} presents the observed brightness temperature spectrum and size of quiet Sun on 2021-08-07 and 2021-08-14 respectively.

\begin{center}
\begin{longtable}{l|c|c|c|c|c} 
\caption{Disk center $T_b$,  {radius(Arcsec) at 0.1~MK ($R_{\rm 0.1MK}$), and from Gaussian fit ($R_{\rm fit} $)} on 2021-08-07, the brightness temperature is an average of the region of $r<0.5R_{sun}$, the error range represents the standard deviation of the values within the region. The error range of the radius represents the {projected beamsize}.
\label{long}}\\
 \hline \multicolumn{6}{| c |}{2021-08-07}\\
 \hline
 {$f$(MHz)}  & $T_B$ ($10^3$ K) & $R_{0.1\rm MK}$ EW  & $R_{0.1\rm MK}$ NS & $R_{\rm fit} $ EW  & $R_{\rm fit} $ NS \\
 \hline \endfirsthead

 \hline
 \multicolumn{6}{|c|}{Continuation of Table \ref{long}}\\
 \hline
 {$f$(MHz)}  & $T_B$ ($10^3$ K) & $R_{0.1\rm MK}$ EW  & $R_{0.1\rm MK}$ NS & $R_{\rm fit} $ EW  & $R_{\rm fit} $ NS \\
 \hline\endhead

 \hline\endfoot
 \hline\hline\endlastfoot
 \hline
    21.48 & 376.9 $\pm$ 6.8 & 2544 $\pm$ 266 & 2152 $\pm$ 218 & 1947 $\pm$ 199 & 1644 $\pm$ 199 \\
   23.63 & 383.2 $\pm$ 9.6 & 2304 $\pm$ 223 & 1952 $\pm$ 182 & 1736 $\pm$ 75 & 1473 $\pm$ 75 \\
   24.6 & 386.9 $\pm$ 8.6 & 2280 $\pm$ 201 & 1928 $\pm$ 159 & 1706 $\pm$ 73 & 1438 $\pm$ 73 \\
   25.58 & 390.4 $\pm$ 9.1 & 2248 $\pm$ 184 & 1880 $\pm$ 140 & 1679 $\pm$ 56 & 1381 $\pm$ 56 \\
   26.56 & 371.9 $\pm$ 10.3 & 2184 $\pm$ 265 & 1792 $\pm$ 225 & 1696 $\pm$ 76 & 1313 $\pm$ 76 \\
   27.53 & 402.3 $\pm$ 27.3 & 2368 $\pm$ 245 & 1768 $\pm$ 206 & 1718 $\pm$ 136 & 1255 $\pm$ 136 \\
   28.51 & 418.6 $\pm$ 13.4 & 2216 $\pm$ 237 & 1800 $\pm$ 198 & 1582 $\pm$ 54 & 1286 $\pm$ 54 \\
   29.29 & 428.9 $\pm$ 14.0 & 2136 $\pm$ 150 & 1784 $\pm$ 65 & 1527 $\pm$ 91 & 1294 $\pm$ 91 \\
   30.46 & 445.0 $\pm$ 14.3 & 2152 $\pm$ 142 & 1768 $\pm$ 56 & 1519 $\pm$ 86 & 1248 $\pm$ 86 \\
   31.44 & 458.2 $\pm$ 17.3 & 2144 $\pm$ 134 & 1712 $\pm$ 51 & 1492 $\pm$ 81 & 1208 $\pm$ 81 \\
   33.59 & 482.1 $\pm$ 20.7 & 2080 $\pm$ 120 & 1680 $\pm$ 43 & 1437 $\pm$ 81 & 1176 $\pm$ 81 \\
   34.76 & 497.8 $\pm$ 23.2 & 2064 $\pm$ 113 & 1664 $\pm$ 40 & 1407 $\pm$ 74 & 1153 $\pm$ 74 \\
   35.74 & 502.8 $\pm$ 31.0 & 2032 $\pm$ 109 & 1688 $\pm$ 37 & 1405 $\pm$ 73 & 1138 $\pm$ 73 \\
   36.91 & 520.1 $\pm$ 28.8 & 2016 $\pm$ 104 & 1640 $\pm$ 36 & 1371 $\pm$ 85 & 1124 $\pm$ 85 \\
   37.88 & 531.4 $\pm$ 32.5 & 1992 $\pm$ 100 & 1624 $\pm$ 35 & 1347 $\pm$ 77 & 1118 $\pm$ 77 \\
   39.25 & 539.7 $\pm$ 29.0 & 1992 $\pm$ 144 & 1616 $\pm$ 113 & 1356 $\pm$ 61 & 1096 $\pm$ 61 \\
   40.23 & 545.8 $\pm$ 30.9 & 1992 $\pm$ 139 & 1608 $\pm$ 110 & 1340 $\pm$ 63 & 1082 $\pm$ 63 \\
   41.2 & 563.6 $\pm$ 45.3 & 1952 $\pm$ 87 & 1584 $\pm$ 31 & 1324 $\pm$ 100 & 1085 $\pm$ 100 \\
   42.18 & 571.3 $\pm$ 44.9 & 1960 $\pm$ 84 & 1568 $\pm$ 30 & 1306 $\pm$ 76 & 1070 $\pm$ 76 \\
   43.16 & 586.4 $\pm$ 48.2 & 1968 $\pm$ 81 & 1560 $\pm$ 29 & 1295 $\pm$ 79 & 1049 $\pm$ 79 \\
   45.31 & 601.4 $\pm$ 53.3 & 1952 $\pm$ 73 & 1520 $\pm$ 27 & 1271 $\pm$ 84 & 1027 $\pm$ 84 \\
   46.28 & 611.1 $\pm$ 56.1 & 1952 $\pm$ 69 & 1512 $\pm$ 26 & 1266 $\pm$ 85 & 1030 $\pm$ 85 \\
   47.65 & 612.6 $\pm$ 44.0 & 1928 $\pm$ 118 & 1520 $\pm$ 92 & 1275 $\pm$ 63 & 1008 $\pm$ 63 \\
   48.63 & 626.5 $\pm$ 59.6 & 1944 $\pm$ 62 & 1488 $\pm$ 24 & 1254 $\pm$ 82 & 997 $\pm$ 82 \\
   49.8 & 629.8 $\pm$ 48.1 & 1912 $\pm$ 110 & 1504 $\pm$ 86 & 1258 $\pm$ 62 & 990 $\pm$ 62 \\
   51.36 & 637.4 $\pm$ 49.3 & 1896 $\pm$ 105 & 1488 $\pm$ 81 & 1250 $\pm$ 64 & 981 $\pm$ 64 \\
   52.53 & 646.2 $\pm$ 51.5 & 1872 $\pm$ 101 & 1472 $\pm$ 75 & 1237 $\pm$ 66 & 971 $\pm$ 66 \\
   53.51 & 654.4 $\pm$ 52.7 & 1872 $\pm$ 98 & 1472 $\pm$ 68 & 1235 $\pm$ 64 & 971 $\pm$ 64 \\
   58.59 & 669.9 $\pm$ 66.4 & 1808 $\pm$ 38 & 1384 $\pm$ 18 & 1211 $\pm$ 114 & 915 $\pm$ 114 \\
   58.98 & 671.0 $\pm$ 55.2 & 1808 $\pm$ 94 & 1424 $\pm$ 71 & 1203 $\pm$ 64 & 920 $\pm$ 64 \\
   60.93 & 695.7 $\pm$ 56.5 & 1792 $\pm$ 95 & 1408 $\pm$ 73 & 1194 $\pm$ 64 & 908 $\pm$ 64 \\
   61.71 & 764.8 $\pm$ 74.6 & 1776 $\pm$ 28 & 1368 $\pm$ 17 & 1203 $\pm$ 142 & 898 $\pm$ 142 \\
   64.06 & 755.1 $\pm$ 79.0 & 1776 $\pm$ 25 & 1344 $\pm$ 16 & 1227 $\pm$ 172 & 891 $\pm$ 172 \\
   64.45 & 767.4 $\pm$ 82.3 & 1760 $\pm$ 24 & 1344 $\pm$ 16 & 1222 $\pm$ 173 & 891 $\pm$ 173 \\
   66.01 & 760.3 $\pm$ 86.6 & 1744 $\pm$ 22 & 1336 $\pm$ 15 & 1188 $\pm$ 195 & 896 $\pm$ 195 \\
   66.4 & 783.0 $\pm$ 66.0 & 1760 $\pm$ 85 & 1400 $\pm$ 63 & 1167 $\pm$ 62 & 885 $\pm$ 62 \\
   69.53 & 767.5 $\pm$ 64.9 & 1728 $\pm$ 80 & 1376 $\pm$ 59 & 1152 $\pm$ 60 & 873 $\pm$ 60 \\
   70.31 & 733.1 $\pm$ 90.7 & 1696 $\pm$ 20 & 1304 $\pm$ 14 & 1151 $\pm$ 243 & 893 $\pm$ 243 \\
   71.48 & 730.4 $\pm$ 91.7 & 1672 $\pm$ 19 & 1296 $\pm$ 13 & 1154 $\pm$ 246 & 882 $\pm$ 246 \\
   72.26 & 746.0 $\pm$ 62.8 & 1696 $\pm$ 81 & 1360 $\pm$ 39 & 1150 $\pm$ 74 & 868 $\pm$ 74 \\
   73.43 & 726.9 $\pm$ 86.9 & 1664 $\pm$ 18 & 1296 $\pm$ 13 & 1155 $\pm$ 242 & 860 $\pm$ 242 \\
   74.21 & 746.8 $\pm$ 63.1 & 1688 $\pm$ 80 & 1352 $\pm$ 36 & 1136 $\pm$ 72 & 858 $\pm$ 72 \\
   76.95 & 748.9 $\pm$ 64.6 & 1696 $\pm$ 77 & 1336 $\pm$ 34 & 1129 $\pm$ 67 & 845 $\pm$ 67 \\
   78.12 & 746.7 $\pm$ 65.1 & 1664 $\pm$ 75 & 1312 $\pm$ 33 & 1125 $\pm$ 70 & 839 $\pm$ 70 \\
   78.9 & 749.8 $\pm$ 125.1 & 1608 $\pm$ 16 & 1264 $\pm$ 12 & 1108 $\pm$ 206 & 831 $\pm$ 206 \\
   80.07 & 737.3 $\pm$ 76.2 & 1576 $\pm$ 74 & 1336 $\pm$ 32 & 1088 $\pm$ 106 & 845 $\pm$ 106 \\
 \hline
\end{longtable}
\end{center}

\begin{center}
\begin{longtable}{l|c|c|c|c|c} 
\caption{Disk center $T_b$,  {radius(Arcsec) at 0.1~MK ($R_{\rm 0.1MK}$), and from Gaussian fit ($R_{\rm fit} $)} on 2021-08-14, the brightness temperature is  an average of the region of $r<0.5R_{sun}$, the error range represents the standard deviation of the values within the region. The error range of the radius represents the {projected beamsize}.
\label{long2}}\\
 \hline \multicolumn{6}{| c |}{2021-08-14}\\
 \hline
 {$f$(MHz)}  & $T_B$ ($10^3$ K) & $R_{0.1\rm MK}$ EW  & $R_{0.1\rm MK}$ NS & $R_{\rm fit} $ EW  & $R_{\rm fit} $ NS \\
 \hline \endfirsthead

 \hline
 \multicolumn{6}{|c|}{Continuation of Table \ref{long2}}\\
 \hline
 {$f$(MHz)}  & $T_B$ ($10^3$ K) & $R_{0.1\rm MK}$ EW  & $R_{0.1\rm MK}$ NS & $R_{\rm fit} $ EW  & $R_{\rm fit} $ NS \\
 \hline\endhead

 \hline\endfoot
 \hline\hline\endlastfoot

 \hline
 21.48 & 424.8 $\pm$ 12.5 & 2840 $\pm$ 284 & 2032 $\pm$ 221 & 2120 $\pm$ 25 & 1460 $\pm$ 25 \\
   23.63 & 429.5 $\pm$ 14.3 & 2536 $\pm$ 240 & 1976 $\pm$ 190 & 1930 $\pm$ 58 & 1414 $\pm$ 58 \\
   24.6 & 446.9 $\pm$ 21.3 & 2504 $\pm$ 223 & 1888 $\pm$ 175 & 1875 $\pm$ 40 & 1328 $\pm$ 40 \\
   25.58 & 457.3 $\pm$ 19.4 & 2496 $\pm$ 210 & 1864 $\pm$ 164 & 1863 $\pm$ 63 & 1308 $\pm$ 63 \\
   29.29 & 515.9 $\pm$ 36.4 & 2384 $\pm$ 160 & 1784 $\pm$ 108 & 1752 $\pm$ 100 & 1233 $\pm$ 100 \\
   30.46 & 532.0 $\pm$ 38.4 & 2384 $\pm$ 149 & 1728 $\pm$ 88 & 1765 $\pm$ 111 & 1202 $\pm$ 111 \\
   31.44 & 527.3 $\pm$ 41.8 & 2336 $\pm$ 154 & 1696 $\pm$ 57 & 1704 $\pm$ 115 & 1177 $\pm$ 115 \\
   33.59 & 573.4 $\pm$ 47.7 & 2288 $\pm$ 138 & 1704 $\pm$ 42 & 1666 $\pm$ 99 & 1169 $\pm$ 99 \\
   34.76 & 590.2 $\pm$ 51.1 & 2256 $\pm$ 131 & 1664 $\pm$ 38 & 1627 $\pm$ 88 & 1147 $\pm$ 88 \\
   35.74 & 582.2 $\pm$ 52.0 & 2264 $\pm$ 128 & 1624 $\pm$ 44 & 1603 $\pm$ 72 & 1137 $\pm$ 72 \\
   36.91 & 618.2 $\pm$ 58.6 & 2208 $\pm$ 121 & 1656 $\pm$ 34 & 1546 $\pm$ 82 & 1128 $\pm$ 82 \\
   37.88 & 629.6 $\pm$ 61.5 & 2192 $\pm$ 117 & 1640 $\pm$ 34 & 1534 $\pm$ 86 & 1117 $\pm$ 86 \\
   39.25 & 642.8 $\pm$ 56.6 & 2192 $\pm$ 158 & 1624 $\pm$ 125 & 1518 $\pm$ 64 & 1094 $\pm$ 64 \\
   40.23 & 652.6 $\pm$ 58.1 & 2176 $\pm$ 152 & 1624 $\pm$ 120 & 1496 $\pm$ 62 & 1084 $\pm$ 62 \\
   41.2 & 666.2 $\pm$ 70.8 & 2160 $\pm$ 98 & 1576 $\pm$ 28 & 1464 $\pm$ 67 & 1069 $\pm$ 67 \\
   42.18 & 681.4 $\pm$ 71.7 & 2120 $\pm$ 101 & 1584 $\pm$ 29 & 1452 $\pm$ 73 & 1065 $\pm$ 73 \\
   43.16 & 691.4 $\pm$ 74.0 & 2096 $\pm$ 96 & 1584 $\pm$ 28 & 1436 $\pm$ 69 & 1059 $\pm$ 69 \\
   45.31 & 710.8 $\pm$ 82.7 & 2064 $\pm$ 87 & 1560 $\pm$ 26 & 1397 $\pm$ 68 & 1033 $\pm$ 68 \\
   46.28 & 711.3 $\pm$ 91.8 & 2064 $\pm$ 70 & 1544 $\pm$ 18 & 1371 $\pm$ 61 & 1030 $\pm$ 61 \\
   47.65 & 725.9 $\pm$ 72.9 & 2040 $\pm$ 127 & 1544 $\pm$ 98 & 1377 $\pm$ 55 & 1014 $\pm$ 55 \\
   48.63 & 739.1 $\pm$ 89.8 & 2016 $\pm$ 73 & 1536 $\pm$ 22 & 1351 $\pm$ 63 & 1012 $\pm$ 63 \\
   49.8 & 741.6 $\pm$ 78.6 & 2016 $\pm$ 119 & 1528 $\pm$ 92 & 1346 $\pm$ 50 & 1003 $\pm$ 50 \\
   51.36 & 756.6 $\pm$ 81.5 & 1984 $\pm$ 113 & 1504 $\pm$ 87 & 1326 $\pm$ 46 & 987 $\pm$ 46 \\
   52.53 & 762.8 $\pm$ 79.9 & 1968 $\pm$ 108 & 1496 $\pm$ 81 & 1312 $\pm$ 46 & 984 $\pm$ 46 \\
   53.51 & 764.1 $\pm$ 83.3 & 1952 $\pm$ 105 & 1480 $\pm$ 80 & 1302 $\pm$ 45 & 978 $\pm$ 45 \\
   58.59 & 781.2 $\pm$ 106.2 & 1880 $\pm$ 45 & 1432 $\pm$ 16 & 1243 $\pm$ 48 & 948 $\pm$ 48 \\
   58.98 & 778.9 $\pm$ 86.5 & 1888 $\pm$ 94 & 1440 $\pm$ 67 & 1247 $\pm$ 36 & 943 $\pm$ 36 \\
   60.93 & 815.8 $\pm$ 91.6 & 1880 $\pm$ 91 & 1424 $\pm$ 65 & 1230 $\pm$ 34 & 924 $\pm$ 34 \\
   61.71 & 892.6 $\pm$ 137.8 & 1856 $\pm$ 28 & 1424 $\pm$ 17 & 1225 $\pm$ 62 & 930 $\pm$ 62 \\
   64.06 & 878.6 $\pm$ 145.6 & 1824 $\pm$ 25 & 1408 $\pm$ 16 & 1236 $\pm$ 80 & 913 $\pm$ 80 \\
   64.45 & 882.9 $\pm$ 150.9 & 1840 $\pm$ 25 & 1392 $\pm$ 15 & 1247 $\pm$ 84 & 912 $\pm$ 84 \\
   66.01 & 865.9 $\pm$ 154.8 & 1824 $\pm$ 22 & 1384 $\pm$ 15 & 1263 $\pm$ 90 & 904 $\pm$ 90 \\
   66.4 & 903.5 $\pm$ 106.7 & 1864 $\pm$ 82 & 1408 $\pm$ 55 & 1203 $\pm$ 32 & 908 $\pm$ 32 \\
   69.53 & 880.2 $\pm$ 108.5 & 1832 $\pm$ 81 & 1392 $\pm$ 56 & 1187 $\pm$ 31 & 897 $\pm$ 31 \\
   70.31 & 815.1 $\pm$ 162.4 & 1800 $\pm$ 20 & 1336 $\pm$ 13 & 1235 $\pm$ 125 & 891 $\pm$ 125 \\
   71.48 & 814.2 $\pm$ 162.7 & 1776 $\pm$ 19 & 1344 $\pm$ 12 & 1221 $\pm$ 121 & 895 $\pm$ 121 \\
   72.26 & 863.1 $\pm$ 110.3 & 1808 $\pm$ 84 & 1368 $\pm$ 32 & 1180 $\pm$ 37 & 888 $\pm$ 37 \\
   73.43 & 815.8 $\pm$ 156.4 & 1744 $\pm$ 18 & 1328 $\pm$ 12 & 1221 $\pm$ 129 & 880 $\pm$ 129 \\
   74.21 & 854.0 $\pm$ 107.2 & 1784 $\pm$ 80 & 1368 $\pm$ 30 & 1170 $\pm$ 40 & 887 $\pm$ 40 \\
   76.95 & 846.3 $\pm$ 109.9 & 1768 $\pm$ 76 & 1360 $\pm$ 28 & 1159 $\pm$ 39 & 882 $\pm$ 39 \\
   78.12 & 845.6 $\pm$ 109.8 & 1752 $\pm$ 74 & 1368 $\pm$ 27 & 1159 $\pm$ 40 & 878 $\pm$ 40 \\
   78.9 & 818.6 $\pm$ 153.5 & 1672 $\pm$ 17 & 1272 $\pm$ 12 & 1172 $\pm$ 134 & 881 $\pm$ 134 \\
   80.07 & 730.0 $\pm$ 129.1 & 1704 $\pm$ 71 & 1320 $\pm$ 26 & 1191 $\pm$ 103 & 886 $\pm$ 103 \\
 \hline
\end{longtable}
\end{center}

\bibliography{cite}

\clearpage

%\end{linenumbers}
\end{document}